\documentclass[pre,twocolumn,aps,eqsecnum]{revtex4}
\usepackage{epsfig}
\usepackage{graphicx}
\usepackage{latexsym}

\begin{document}

\title{Steady-state, effective-temperature dynamics in a glassy material}

\author{ J. S. Langer and M. L. Manning}
\affiliation{Dept. of Physics, University of California, Santa Barbara, CA  93106-9530}

\date{\today}

\begin{abstract}
We present an STZ-based analysis of numerical simulations by Haxton and Liu (HL).  The extensive HL data sharply test the basic assumptions of the STZ theory, especially the central role played by the effective disorder temperature as a dynamical state variable.  We find that the theory survives these tests, and that the HL data provide important and interesting constraints on some of its specific ingredients.  Our most surprising conclusion is that, when driven at various constant shear rates in the low-temperature glassy state, the HL system exhibits a classic glass transition, including super-Arrhenius behavior, as a function of the effective temperature.
\end{abstract}
\maketitle

\section{Introduction: Basic Hypotheses}
\label{hypotheses}

In a remarkable recent report, T. Haxton and A. Liu (HL)~\cite{HL07} have described molecular-dynamics simulations of a simple, sheared, two-dimensional, glass-forming material over three decades of steady-state strain rates $\dot\gamma$, and for bath temperatures $T$ ranging from about one tenth of the glass transition temperature $T_0$ to about twice $T_0$.  Most importantly, by measuring pressure fluctuations, HL have determined values of the effective disorder temperature $T_{\mbox{\small eff}}$ for each value of $\dot\gamma$, $T$, and the shear stress $s$.  This extensive data set tests the applicability of any theory of amorphous plasticity such as the shear-transformation-zone (STZ) theory discussed here, and also probes the limits of validity of the effective-temperature concept.  \cite{ONOetal,CUGLIANDOLOetal,SOLLICHetal,BERTHIER-BARRAT,OHERNetal,ILG-BARRAT} We find that the main features of the STZ theory presented in earlier papers~\cite{FL98,JSL04,BLP07I,BLP07II} nicely survive these tests.  Moreover, several of the theory's specific ingredients, which had not been tightly constrained by earlier experiments or simulations, can be refined and extended in light of the HL data.  Our most surprising conclusion is that, for $T < T_0$, the relation between $\dot\gamma$ and $T_{\mbox{\small eff}}$ is a direct analog of the  relation between the $\alpha$ relaxation rate and the bath temperature $T$ near a conventional glass transition.  

\begin{figure}[h]
\centering \includegraphics[height=9.5cm]{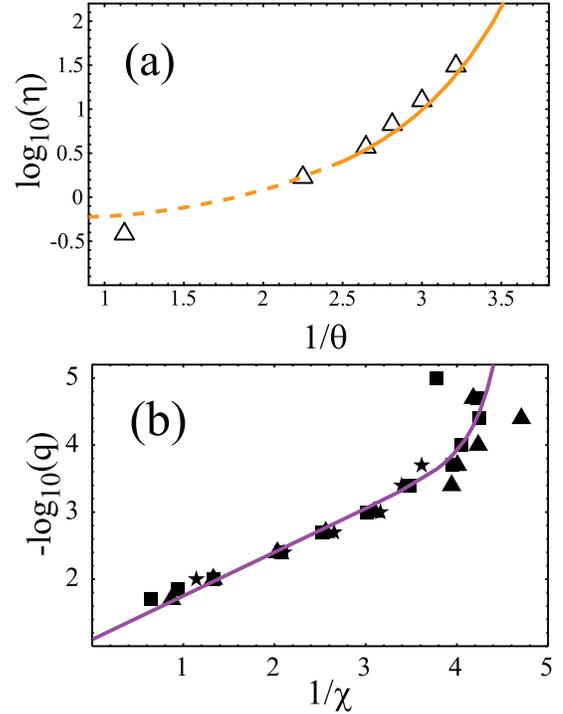}
\caption{\label{Arr1}(a) Logarithm of the HL viscosity (open triangles) as a function of the inverse temperature $1/\theta = T_Z/T$. The solid line is the STZ-theory fit to these data points. The dashed extension of this line indicates a region where the temperatures are so high that the theory becomes incorrect. (b) Logarithm of the HL strain rate as a function of the inverse effective temperature $1/\chi$ at temperatures below the glass transition temperature $\theta_0 = 0.20$. The data points correspond to temperatures $\theta = 0.022$ (stars),$0.044$ (squares) and $0.11$ (triangles). The solid line is the theoretical fit.  See  Section \ref{HLdata} for details of the theoretical analysis and an evaluation of $T_Z$.}
\end{figure}

It is well-known (e.g. see \cite{TARJUS-KIVELSON}) that, for unstressed glass-forming materials in a range of  temperatures above some $T_A$, the viscosity $\eta$ is -- at least to a first approximation -- a simple activated process with a temperature-independent energy barrier. Below $T_A$, $\eta$ depends super-Arrheniusly on $T$, diverging as $T$ decreases toward the glass transition temperature (or  Kauzmann temperature) $T_0$.  Figure~\ref{Arr1}(a) shows that the HL data for the viscosity in the limit of vanishing strain rate can be fit by a function of the form
\begin{equation}
\label{eta-T}
\eta = \eta_0(\theta) \exp \left[{1\over\theta} + \alpha(\theta)\right], 
\end{equation}
where $\theta = T/T_Z$ and $k_B\,T_Z$ is an Arrhenius activation energy that we interpret as the energy of formation for an STZ.   $\alpha(\theta)$ is a model-dependent function, to be specified later, that diverges as $\theta \to \theta_0 = T_0/T_Z$ and vanishes for $\theta > \theta_A = T_A/T_Z$.  The prefactor $\eta_0(\theta)$ is a relatively slowly varying function of $\theta$ that appears prominently in our data analysis in Section \ref{HLdata}.

Our key hypothesis is that, below the glass transition temperature $\theta_0$, there is a relationship analogous to Eq.(\ref{eta-T}) between the steady-state effective temperature and the strain rate $\dot\gamma$.  In analogy to the definition of $\theta$, and consistent with earlier notation \cite{JSL04,BLP07I,BLP07II}, we write $\chi \equiv T_{\mbox{\small eff}}/T_Z$.  We also define $q \equiv \tau_0\,\dot\gamma$, where $\tau_0$ is the fundamental time scale that appears in the dynamical STZ equations of motion.  Thus, $q$ is a dimensionless rate of molecular rearrangements or, equivalently, a stirring rate.  The HL results are expressed in terms of a molecular time scale, $t_0$, which is derived from their molecular interactions and is not necessarily the same as $\tau_0$ -- although the two times should have about the same order of magnitude. We find it convenient to set $\tau_0 = t_0 = 1$, and to incorporate any difference between these time scales into the parameter $\epsilon_0$ defined below in Eq.~(\ref{f-s}).  For clarity, however, we occasionally retain explicit factors $\tau_0$. 

We propose that the rate $q$ be interpreted as a purely kinetic quantity, independent of the specific mechanism by which the rearrangements occur.  We then define $\hat\chi(q)$ to be the dimensionless, steady-state, effective temperature at stirring rate $q$.  To a first approximation, we assume that $\hat\chi(q)$ is independent of $\theta$ for $\theta \le \theta_0$.  As we shall see later, the temperature dependence of $\hat\chi(q)$ becomes more interesting for larger $\theta$.

In the limit $q \ll 1$, the number of rearrangements depends only on the extent of the deformation and not its duration; thus the condition for steady-state equilibrium is that, with high probability, each molecule has changed its neighbors at least a few times.  It follows that $\hat\chi(q \to 0) = \chi_0$ is a nonzero constant, also independent of the rearrangement mechanism.  We then postulate that the inverse function $q(\hat\chi)$ is the rate at which rearrangements occur when the system is driven so that its steady-state effective temperature is $\hat\chi$.  Then, in analogy with Eq.(\ref{eta-T}), we write
\begin{equation}
\label{q-chi}
\frac{1}{q(\hat\chi)} = {1 \over q_0} \exp \left[ {A\over \hat\chi} + \alpha_{\mbox{\small eff}}(\hat\chi)\right],
\end{equation}
where $q_0$ and $A$ are constants to be determined, and $\alpha_{\mbox{\small eff}}(\hat\chi)$ is a super-Arrhenius function that diverges at the effective Kauzmann temperature $\chi_0$ and vanishes above some $\chi_A$. 

This hypothesis is supported by the HL data.  Fig.~\ref{Arr1}(b) is an Arrhenius plot of the HL results for the strain rate versus the effective temperature for three values of $\theta$, all almost certainly below $\theta_0$, and the solid curve is a fit to those points using Eq.~(\ref{q-chi}). There are large uncertainties in the data at  small strain rates (large $1/\chi$), but the existence of Arrhenius and super-Arrhenius regions is apparent.  Note that the Arrhenius region, i.e. the section of the curve that is linear in $1/\chi$, seems to extend all the way to infinite $\chi$.  Ordinarily we do not think of Arrhenius formulas remaining valid out to high temperatures.  The difference here may be that the effective temperature pertains only to a fraction of the degrees of freedom in this system; so, perhaps in analogy to magnetic spin temperatures, the maximally disordered state at infinite $\chi$ is realistic. 

One of the most interesting features of Eq.(\ref{q-chi}) is that $\chi$ diverges at a finite strain rate, $q_0 \cong 0.08$ as seen from the extrapolation shown in Fig.~\ref{Arr1}(b).  For molecular systems where $\tau_0$ is of order femtoseconds, $\dot\gamma = 0.08/\tau_0$ is an impossibly large rate; but it is accessible in numerical simulations and perhaps also in granular systems or the like. The system must liquify in some sense when driven faster than $q_0$; and the STZ theory must break down near that point. 

\section{STZ Theory}
\label{STZeqns}

\subsection{General STZ Equations}

In order to use Eq.(\ref{q-chi}) to compute stresses, strain rates and effective temperatures for $\theta$ both below and above $\theta_0$, we need a complete set of dynamical equations of the kind provided by the STZ theory.  In this Section, we summarize the version of that theory that was  presented in \cite{JSL04}, specializing to steady-state motion but otherwise retaining the generality needed to describe plasticity over the wide range of driving conditions studied by HL.  We need one  structural modification of this theory, which we explain following Eq.(\ref{m-s}) below; and our analysis requires model dependent rate factors that differ from those used in earlier papers.   We start by writing the STZ equations in a form determined by the basic structure of the theory.  Then, in the following subsection, we specify our choice of the model dependent ingredients of these equations.

In keeping with the emphasis of HL and the results of recent STZ analyses such as \cite{BLP07I,BLP07II}, we write this theory in a form in which the effective temperature $\chi$ is the single, dynamically relevant, internal state variable.  Accordingly, we need just one equation for the dimensionless strain rate $q$ as a function of the shear stress $s$ and the effective temperature $\chi$, supplemented by the steady-state version of an equation of motion for $\chi$.  

The equation for the strain rate has the form
\begin{equation}
\label{q-s}
q \equiv \dot\gamma\,\tau_0 =2 \, e^{-1/\chi}\,f(s,\theta),
\end{equation}
where $f(s,\theta)$ is a model-dependent function (to be specified shortly), and $\exp (-1/\chi)$ is proportional to the STZ density. This clean separation between the stress-dependent and $\chi$-dependent parts of the plastic strain rate is an important characteristic feature of the STZ theory that has been emphasized recently by Shi et. al. \cite{SHI06}.  

The equation for $\chi$ is
\begin{equation}
\label{chi1}
e^{-1/\chi}\,\Gamma(s,\theta)\,\left[1-{\chi\over \hat\chi(q)}\right]=K(\chi)\,\rho(\theta)\,\left({\chi\over\theta}-1\right).
\end{equation}
This is essentially the steady-state version of Eq.(3.5) in \cite{JSL04}; but here we have written it in a form that is more appropriate for situations in which $\hat\chi(q)$ can become large. 

According to Eq.(\ref{chi1}), $\chi$ finds its steady-state value when the rate at which it is being driven mechanically toward its  kinetically optimal value $\hat\chi(q)$, on the left-hand side, is balanced by the rate at which $\chi$ relaxes toward the bath temperature on the right. $\Gamma(s,\theta)$, another model-dependent function, is proportional to the rate at which the work done by the driving force is dissipated as configurational entropy; thus $\Gamma(s,\theta)$ vanishes at $s=0$. The function 
\begin{equation}
\label{rhodef}
\rho(\theta)\equiv \cases{\exp\,[-\alpha(\theta)]&for $\theta > \theta_0$\cr 0 & for $\theta < \theta_0$}
\end{equation}
is the super-Arrhenius part of the $\alpha$ relaxation rate.  Because $\rho(\theta)$ vanishes for $\theta < \theta_0$,  Eq.(\ref{chi1}) implies that the low-temperature, steady-state value of $\chi$ is $\hat\chi(q)$ as anticipated in Eq.(\ref{q-chi}). The function $K(\chi)$ appearing in Eq.(\ref{chi1}) is proportional to the density of sites at which thermal equilibration events -- as opposed to shear-transformations -- take place. We assume here, as in \cite{JSL04}, that 
\begin{equation}
\label{K-chi}
K(\chi)= \kappa\,e^{-\beta/\chi}, 
\end{equation}
with $\beta=1$, which means that both kinds of sites are comparably populated.  The parameter $\kappa$ remains to be determined from the data.   

In the general STZ equations as developed in \cite{JSL04,BLP07II},  the function $f(s,\theta)$ in Eq.(\ref{q-s}) has the form 
\begin{equation}
\label{f-s}
f(s,\theta)= \epsilon_0\,{\cal C}(s)\,\Bigl[{\cal T}(s)-m(s)\Bigr],
\end{equation}
where $\epsilon_0$ is a dimensionless constant,
\begin{equation}
{\cal C}(s)={1\over 2}\,\Bigl[R(s)+R(-s)\Bigr],
\end{equation}
\begin{equation}
{\cal T}(s)={R(s)-R(-s)\over R(s)+R(-s)},
\end{equation}
and $R(s)/\tau_0$ is the rate at which individual STZ's undergo forward shear transitions driven by the stress $s$.  

The function $m(s)$ is the orientational variable that emerges in the role of a back stress in the STZ theory.  It satisfies the steady-state relation:
\begin{equation}
\label{m-s}
2\,{\cal C}(s)\,\Bigl[{\cal T}(s)-m(s)\Bigr]\Bigl[1-m(s)\,s/s_0 \Bigr] = m(s)\,\rho(\theta).
\end{equation}
Here, and in the equation for $\Gamma(s,\theta)$ that follows, we depart slightly from the equations derived in \cite{JSL04}.  In that paper, the yield stress $s_y$ was assumed to be approximately equal to a temperature-independent constant $\bar\mu$, and the stress was everywhere written in units of that constant.  Here, on the other hand, it is obvious -- most visibly in the HL graphs of stress versus strain rate at low-temperatures -- that $s_y$ must be a function of $\theta$.  Examination of the derivation of the STZ equations in \cite{JSL04} indicates that the factor of proportionality between $\Gamma(s,\theta)$ and the energy dissipation rate per STZ, which was set equal to $\bar\mu$ there primarily for dimensional reasons, should have been a possibly temperature-dependent stress that we now call $s_0$.  The result of this change is that a factor $s$, which multiplies the strain rate in the expression for the rate at which plastic work is done on the system, is replaced by $s/s_0$, and the explicit $s$ that appears in Eq.(\ref{m-s}) now becomes $s/s_0$.  This change in the analysis removes an ambiguity in the older STZ theories; e.g. see \cite{PECHENIK05}  Details will be published elsewhere. \cite{JSLrev08}

The significance of the quantity $s_0$ is that below the glass transition temperature it is closely related to the dynamical yield stress $s_y$.  For $\theta < \theta_0$ and $\rho(\theta)=0$, $s_y$ is the value of $s$ for which the two factors on the left-hand side of Eq.(\ref{m-s}) vanish simultaneously; therefore it is the solution of the equation
\begin{equation}
\label{sy}
s_y\,{\cal T}(s_y)=s_0.
\end{equation}
Because ${\cal T}(s) \le 1$ in all cases and is approximately equal to unity at low temperatures,  $s_y \cong s_0$. If $s_0$ is a function of $\theta$ in Eq.(\ref{sy}), then $s_y$ also is temperature dependent.
  Although there is no yield stress for $\theta > \theta_0$, $f(s,\theta)$ still changes abruptly at $s \cong s_0$ for $\theta$ not too much larger than $\theta_0$. This implies that at stresses close to $s_0$, the stress vs. strain rate curve exhibits an abrupt change in slope.   

The STZ formula for the dissipation rate $\Gamma(s,\theta)$ appearing in Eq.~(\ref{chi1}) is
\begin{equation}
\label{Gamma}
\Gamma(s,\theta)={2\,{\cal C}(s)\Bigl[{\cal T}(s)- m\Bigr]\Bigl[s/s_0-\xi(m)\Bigr]+ m\,\xi(m)\,\rho(\theta)\over 1- m\,\xi(m)},
\end{equation}
where $m = m(s)$ as given by Eq.(\ref{m-s}), and 
\begin{equation}
\label{xidef}
s_0\,\xi(m) = {\cal T}^{-1}(m)
\end{equation}
is the inverse of the function ${\cal T}(s)$; that is, ${\cal T}[s_0\,\xi(m)]=m$.  It is not difficult to show that $\epsilon_0\,\Gamma(s,\theta) = 2\,s\,f(s,\theta)$ for $\theta < \theta_0$, confirming that the energy dissipated is equal to the work done on the system when configurational relaxation cannot occur via thermal fluctuations alone.  Above the glass transition, on the other hand, the external work is not the same as the dissipation, and the  use of Eq.(\ref{Gamma}) is necessary for accurate calculations. 

\subsection{Model-Dependent Ingredients}

We turn now to the STZ forward transition rate $R(s)$.  Here we depart from the choices made in earlier papers \cite{JSL04,BLP07II}, where we were concerned with experiments and simulations performed only at relatively small driving forces, and which therefore did not sharply constrain our choice of $R(s)$.  HL have provided a data set that goes up to stresses more than ten times the yield stress, at temperatures both well above and well below the glass transition, and thus have made it necessary for us to consider this rate factor in a broader context than before. 

We  need a function $R(s)$ that interpolates between Eyring-like behavior at very small stresses and power-law growth at very large ones. In choosing such an $R(s)$, we find that we must depart from our earlier, purely phenomenological procedure of choosing simple functions with very few adjustable parameters, in the expectation that whatever data we had available would not justify additional theoretical complications.  Here we are facing a very different situation.  In order to interpret the HL data, we find it better to start with a physically motivated model containing an uncomfortably large number of parameters.  Determining the values of these parameters is made easier in places by the fact that we can fit some parts of the data independently of other parts, and then see how these initially separate pieces of the puzzle fit together, for example, in connecting our predictions of behaviors at very small strain rates to those at very large ones.  Another advantage of this more nearly first-principles procedure is that we can guess the magnitudes of many parameters on physical grounds, and also know the limits of validity of physics-based approximations.  For example, we know that the Eyring-like formula that we use for the STZ transition rate breaks down at the highest temperatures simulated by HL; and thus we know to assign less weight to our theory at those temperatures in fitting the HL viscosity curve.  

Our proposed $R(s)$ is
\begin{equation}
\label{R-s}
R(s) = \exp \left[-{\theta_E\over \theta}\,e^{-s/\tilde\mu}\right]\left[1 + \left({s\over s_1}\right)^{2} \right]^{n/2}.
\end{equation}
The first factor on the right-hand side of Eq.(\ref{R-s}) is the Eyring rate in a form similar to that used in \cite{FL98}, where the exponential function of $s/\tilde\mu$ causes the rate to saturate at large $s$.  Here, $\theta_E$ is the height of the Eyring activation barrier in units of $T_Z$. We expect $\theta_E < 1$, because the barrier opposing STZ transitions from one state to the other should be less than the barrier opposing creation of new STZ's.  The parameter $\tilde\mu$ is related to the curvature of this barrier at its peak.  It appears here ostensibly in the same place where $\bar\mu$ appeared in \cite{JSL04}, but now it has no direct connection to the yield stress. In the limit of small stress, we recover an Eyring rate factor in the form 
\begin{equation}
R(s) \approx \exp \left[-{\theta_E\over \theta}\,(1 - s/\tilde\mu)\right];
\end{equation}
therefore
\begin{equation}
\label{Csmall}
{\cal C}(s) \approx e^{-\theta_E/\theta} \cosh\left({\theta_E\,s\over \theta\,\tilde\mu}\right);
\end{equation}
and
\begin{equation}
\label{Tsmall}
{\cal T}(s) \approx \tanh \left({\theta_E\,s\over \theta\,\tilde\mu}\right).
\end{equation}

The second factor on the right-hand side of Eq.(\ref{R-s}) converts the saturated Eyring function at large $s$ to a power law:
\begin{equation}
\label{Csn}
{\cal C}(s) \approx \left({s\over s_1}\right)^n;~~~~{\cal T}(s) \approx 1.
\end{equation}
Here, $s_1$ is a temperature independent stress scale.  Without loss of generality, we normalize the nominal yield stress $s_0(\theta)$ so that $s_0(0)=1$ and measure stresses $s$ in units of this zero-temperature yield stress. The crossover stress $s_1$ determines when the system departs from Eyering stress-activated behavior.

The exponent $n$ in Eq.(\ref{R-s}) is especially interesting.  In strongly dissipative systems, we expect $n = 1$, indicating that the STZ transition rate is controlled at large stresses by some linear dissipative mechanism such as friction or viscosity.  In the opposite limit, where rates are controlled by collisions between primarily hard-core objects and the detailed molecular interactions are relatively  unimportant, we expect to find Bagnold scaling.\cite{BAGNOLD} That is, if there is no natural energy or stress scale in the problem, then dimensional analysis requires that the stress be proportional to the square of a rate.  Since $R(s)/\tau_0$ is the only available quantity with dimensions of inverse time, we must have $R(s) \sim s^{1/2}$ and $n = 1/2$.  It appears that the latter case is more consistent with the HL viscosity data.

Other model-dependent ingredients that we must specify are the super-Arrhenius function $\alpha(\theta)$ and its analog $\alpha_{\mbox{\small eff}}(\hat\chi)$.  Here we adopt formulas motivated by the excitation-chain theory of the glass transition \cite{JSL-XCPRE06,JSL-XCPRL06} proposed recently by one of us. For present purposes, these formulas need not be interpreted as anything more than phenomenological fits to experimental data. We write
\begin{equation}
\label{alpha-T}
\alpha(\theta) = \left({\theta_1\over \theta - \theta_0}\right)^p\,\exp \left[- b\,{\theta-\theta_0\over \theta_A-\theta_0}\right],
\end{equation}
where $b \cong 3$ produces a reasonably accurate description of the transition from an inverse power-law divergence near $\theta_0$ to simple Arrhenius behavior above $\theta_A$ in analyses of experimental data.\cite{JSLrev08}  The excitation-chain theory predicts $p=1$ (Vogel-Fulcher) in three dimensions and $p=2$ in two, but the HL data is not accurate enough to distinguish between these possibilities.  Since HL fit their two dimensional viscosity data with a Vogel-Fulcher function, we have chosen to be conservative and use $p=1$ in our analyses.  However, we point out in Section \ref{HLdata} that $p=2$ remains an interesting possibility. 

The effective-temperature analog of Eq.(\ref{alpha-T}) is
\begin{equation}
\label{alpha-chi}
\alpha_{\mbox{\small eff}}(\hat\chi)=\left({\chi_1\over \hat\chi-\chi_0}\right)^p\,\exp \left[- b\,{\hat\chi-\chi_0\over \chi_A-\chi_0}\right].
\end{equation}

\section{Comparisons between STZ theory and the HL simulations}
\label{HLdata}

\subsection{Evaluating the Parameters}

We turn now to a detailed numerical comparison between these theoretical predictions and the HL data. We start by deducing values of the parameters $T_Z$ and $A$ in Eq.(\ref{q-chi}) and the exponent $n$ in Eqs.(\ref{R-s}) and (\ref{Csn}), using primarily the data in the extreme Arrhenius region at low temperatures $\theta$ and large values of $\chi$.  

\begin{figure}[h]
\centering \includegraphics[height=5 cm]{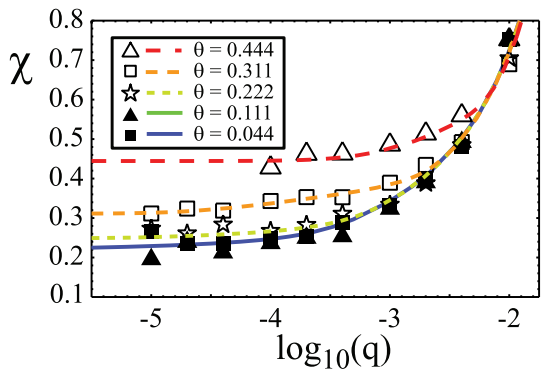}
\caption{\label{HL:qchi} Effective temperature $\chi$ as a function of $\log_{10}(q)$ for HL results both above (open symbols) and below (filled-in symbols) the glass transition. The smooth curves are the predictions of the STZ theory at the same temperatures $\theta$. }
\end{figure}

Figure~\ref{Arr1}(b) shows the HL  data for $-\log_{10}(q) $ as a function of $1/\chi$, for the three lowest HL temperatures, $\theta = 0.022,\,0.044,\,{\rm and}\,\,0.11$, all of which are comfortably below the glass transition temperature $\theta_0\cong 0.20$.  Therefore, $\rho(\theta) = 0$ and $\chi = \hat\chi$.  From Eq.(\ref{q-chi}) and the observed slope of this graph in the Arrhenius region, i.e. from the data shown in HL Fig. 1b, we find that $A\,T_Z \cong .0068$ (in HL units).

A second piece of information is obtained by looking at the HL results for stress as a function of $1/\chi$, shown here in Fig.~\ref{HL:schi}.  Again, we look only in the large-$\chi$, small $\theta$ limit.  Equations (\ref{q-chi}) and (\ref{q-s}), plus our large-$s$ estimate for the rate factor, combine to give 
\begin{equation}
\label{q-s-linear}
q \approx q_0\,e^{- A/\chi} \approx 2\,\epsilon_0\, e^{-1/\chi}\left({s \over s_1}\right)^n 
\end{equation}
from which we find 
\begin{equation}
\label{s-chi-asymptote}
\ln (s/s_1) \approx {1\over n}\,\ln\,\left({q_0\over 2\,\epsilon_0}\right) - {A-1\over n\,\chi}.
\end{equation}
The observed slope in Fig.~\ref{HL:schi}, i.e. in the original HL figure in which $\ln\,(s)$ is plotted as a function of $1/T_{\mbox{\small eff}}$, tells us that $(A-1)\,T_Z/n \cong .004$.  

A third relation, equivalent to the first two but useful for data analysis, is
\begin{equation}
\label{s-q-asymptote}
\ln (s/s_1)\approx {1\over n}\,\ln\left({\,q_0^{1/A}\over\epsilon_0}\right) + {1\over n}\,\left({A-1\over A}\right)\,\ln (q).
\end{equation}
The observed asymptotic slope of the graph of $\ln (s)$ versus $\ln (q)$ shown in Fig.~\ref{HL:sq} is $(A-1)/n\,A \cong 2/3$.  If we assumed $n=1$, we would find $A \cong 3$ and $T_Z\cong 0.00225$, a value of $T_Z$ that is too small to be consistent with the HL viscosity data. However, $n = 1/2$ (the Bagnold prediction) implies that $A \cong 1.5$ and $T_Z \cong 0.0045$, which fits the viscosity quite well.  

\begin{figure}[h]
\centering \includegraphics[height=5 cm]{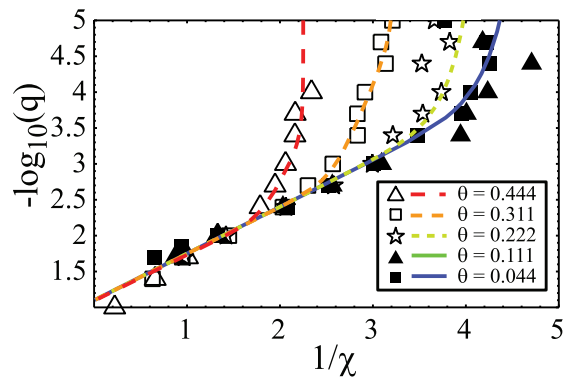}
\caption{\label{HL:qchiall} The same data as shown in Fig.\ref{HL:qchi} but in the form of an Arrhenius plot of $\log_{10}(q)$ as a function of $1/\chi$, for HL results both above (open symbols) and below (filled-in symbols) the glass transition. The smooth curves are the predictions of the STZ theory at the same temperatures $\theta$. }
\end{figure}

We then use the last two graphs to evaluate the $q$-independent terms on the right-hand sides of Eqs.(\ref{s-chi-asymptote}) and (\ref{s-q-asymptote}), with $q_0 \cong .08$. In this way we check for consistency of the estimates we have obtained from a noisy data set, and find a relation between $s_1$ and $\epsilon_0$: 
\begin{equation}
\label{s1}
s_1 \cong 2.3 \times 10^3\,\,\epsilon_0^2. 
\end{equation}

Several other parameters can be determined by direct examination of the HL data.  HL measure their stresses $\sigma$ in units related to their interatomic forces. Because our values of $s$ are expressed in units of the zero-temperature yield stress, the ratio of $\sigma$ to $s$ is approximately the value of $\sigma$ at the lowest strain rate and the lowest temperature shown in  HL Fig.1a, i.e. approximately $0.001$ in HL stress units. Thus, the stress-conversion relation is
\begin{equation}
\label{sigmaHL}
\sigma \cong \sigma_{HL}\,s; ~~~~~\sigma_{HL} \cong 0.001.
\end{equation}  
Similarly, we deduce from the small-$q$ limits of the stress in HL Fig.1a (shown here in Fig.~\ref{HL:sq})  that, for $\theta < \theta_0$, $s_y \cong s_0$ is roughly a linearly decreasing function of $\theta$.  There is no comparably systematic way to evaluate $s_0(\theta)$ for $\theta > \theta_0$; but we see no structure in these curves that might indicate a further decrease of the stress scale $s_0(\theta)$ at temperatures above $\theta_0$.  Accordingly, we choose
\begin{equation}
\label{s0}
s_0(\theta)\cong \cases{1 - c\,\theta/\theta_0 &for $\theta < \theta_0$ \cr 1-c & for $\theta > \theta_0$,}
\end{equation}
where $c \cong 0.6$. 

The parameter $\epsilon_0$, in our notation, is proportional to the number density of STZ's in the limit of infinite $\chi$, modified by a time-scale conversion factor of order unity.  We have found no independent estimate of $\epsilon_0$
comparable to the preceding estimates of $T_Z$, $A$, $n$, etc.; but we expect it to be small if the STZ theory is to remain valid at large values of $\chi$.  Eq.~(\ref{s1}) provides an additional constraint. As discussed in the next section, the stress crossover behavior for the full set of HL data provides an estimate of $s_1 \cong 0.08$, and this implies that $\epsilon_0 \cong .006$. 

\begin{figure}[h]
\centering \includegraphics[height=5 cm]{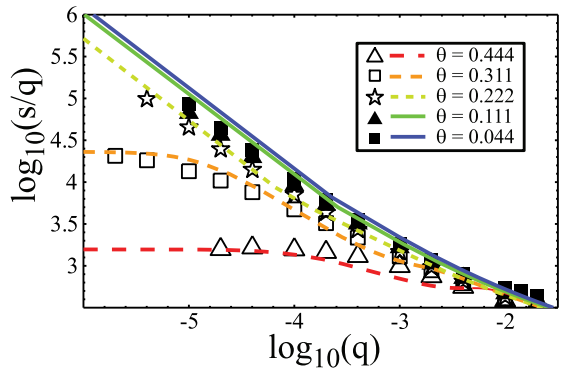}
\caption{\label{HL:etaq} Logarithm of the stress $s$ divided by $q$ as a function of $\log(q)$ for HL results at temperatures both above (open symbols) and below (filled-in symbols) the glass transition. The curves are the predictions of the STZ theory at the same temperatures. For temperatures below the glass transition, the limit as $s \rightarrow 0$ of $s/q$ is the Newtonian viscosity, $\eta$.}
\end{figure}

\begin{figure}[h]
\centering \includegraphics[height=5 cm]{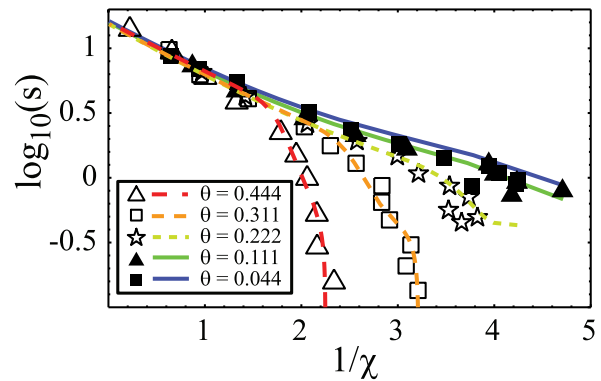}
\caption{\label{HL:schi} Logarithm of the stress $s$ as a function of $1/\chi$ for HL results at temperatures both above (open symbols)  and below (filled-in symbols) the glass transition. The curves are the predictions of the STZ theory at the same temperatures.}
\end{figure}

Our next step is to use the parameters determined so far to compare the HL viscosity shown in Fig.~\ref{Arr1}a with the STZ predictions and thereby evaluate the parameters that occur in Eqs. (\ref{Csmall}), (\ref{Tsmall}), and (\ref{alpha-T}) . Figure~\ref{HL:etaq} shows the HL data for $\log(s/q)$ as a function of $\log(q)$ at temperatures above and below the glass transition. By definition, the HL viscosity is  $\eta = \sigma_{HL}\,s/\dot\gamma$ in the limit $s \to 0$, and the HL viscosity data in Fig.~\ref{Arr1}a are based on extrapolation of the data in Fig.~\ref{HL:etaq} to $s=0$ using the Cross equation.
 
 To evaluate $\eta$, we use Eqs.(\ref{q-s}), (\ref{f-s}), and (\ref{m-s}) with $\chi= \theta$, and, after some algebra, find that
\begin{eqnarray}
\label{eta0}
\eta_0(\theta)&=& {\sigma_{HL}\over \epsilon_0 \,{\cal T}'(0)}\,\left[1+ {\rho(\theta)\over 2\,{\cal C}(0)}\right]\cr &=& {\sigma_{HL}\,\tilde\mu\,\theta\over \epsilon_0\,\theta_E}\,\,\left[1 + {1\over 2}\rho(\theta)\,e^{\theta_E/\theta}\right].
\end{eqnarray}
The second version of this equation makes explicit use of Eqs.(\ref{Csmall}) and (\ref{Tsmall}). Obviously there are too many parameters here for us to obtain more than a very rough estimate of  their range of values.  None of the six HL data points for $\eta$ are close enough to the transition temperature $\theta_0$ to provide an accurate picture of the super-Arrhenius singularity.  To make matters worse, the highest reported temperature is approximately equal to $T_Z$, well above any reasonable value for $\theta_E$ and thus far too high for the Eyring formula to be valid.  Nevertheless, we find some interesting information by exploring the ways in which we can fit the theory to the data.  

Our procedure has been simply to use Eq.(\ref{eta0}) in conjunction with Eq.(\ref{eta-T}) to plot the logarithm of $\eta(\theta)$ and compare the results with the HL data points.  To start, we accept the values $T_Z \cong 0.0045$ (in HL units), obtained from the large-stress behavior, and $\theta_0 \cong 0.20$ estimated by HL.  We also fix $p=1$ and $b =3$ in Eq.(\ref{alpha-T}).  Then our  best-fit parameters are $\theta_A \cong 0.69$, $\theta_1 \cong 0.69$, $\tilde\mu \cong 0.3$, and $\theta_E \cong 0.4$.  Our fit to the HL viscosity measurements is shown in Fig.~\ref{Arr1}(a). The dashed part of that curve indicates the region where, given the above value of $\theta_E$, we know that the Eyring formula in Eq.(\ref{R-s}) cannot be valid.  

In contrast to the uncertainties that complicate the theoretical fit to the HL viscosity data, the analogous relation between the strain rate and the effective temperature shown in Fig.~\ref{Arr1}(b) seems relatively easy to understand.  Whereas the viscosity is an intrinsically dynamic quantity, involving material-specific relations between stress and plastic response, the function $q(\chi)$ appears to be, as we have postulated, a purely kinetic relationship. The solid curve shown in Fig.~\ref{Arr1}(b) is a fit to Eq.(\ref{q-chi}) with the following  parameters defined in Eq.~(\ref{alpha-chi}); $\chi_0 = 0.20$, $\chi_A= 0.30$, $\chi_1 = 0.26$, $p=1$, and $b=3$.  We emphasize that there are still substantial uncertainties in these numbers. Within these uncertainities, it appears that $\chi_0 \cong \theta_0$. 

At the outset, it might seem plausible to set $\theta_0= \chi_0$ and thereby assume that there is only one energy scale that determines both the thermal and effective glass temperatures.  This simplification cannot always be correct, however, because we know from the ubiquitous appearance of transient stress peaks in stress-strain curves measured at high strain rates that slow thermal quenches can bring glasses into states where $\theta_0 < \chi < \chi_0$.  The conventional interpretation of these stress peaks is that the as-quenched state of the system has a low  fictive temperature $\chi$ with a correspondingly small population of STZ's, and that plastic flow cannot begin until enough deformation has occurred to increase $\chi$ to a value of order $\chi_0$.  The fictive temperature cannot be below $\theta_0$, so stress peaks could not be produced by this process if $\theta_0= \chi_0$. In this situation, therefore, we expect that the HL model would not exhibit stress peaks except, possibly, at very high strain rates where $\hat\chi(q)$ is substantially larger than $\chi_0$.  

In exploring fits to the viscosity data, however, we find one intriguingly different possibility.  We  can fit the HL viscosity data with $p=2$  in  Eq.(\ref{alpha-T}) -- the prediction of the excitation-chain theory -- but this fit looks best if we reduce $\theta_0$ to about $0.1$, i.e. half the previous estimate. This lowered transition temperature would mean, for example, that the HL curves of stress versus strain rate shown in Fig.\ref{HL:sq} for values of $\theta$ down to $0.1$, would bend over and exhibit viscous behavior at strain rates much smaller than those measured by HL.  If that were the case, then $\chi_0$ would be substantially higher than $\theta_0$, and the transient behavior of the HL model might be more interesting than predicted at the end of the last paragraph.

\subsection{Comparisons over the Full Range of the HL Data}

In order to extend our analysis to temperatures above the glass transition (Kauzmann) temperature $\theta_0$ and to arbitrary strain rates, we must reexamine our assumption that the effective Kauzmann temperature $\chi_0$ is independent of the bath temperature $\theta$.  To see what is happening here, we take advantage of our simplifying assumption that $\beta = 1$ in Eq.(\ref{K-chi})  and solve this equation for $\chi$:
\begin{equation}
\label{chiss}
{1\over\chi} = {\Gamma(s,\theta)/\hat\chi(q) + \kappa\,\rho(\theta)/\theta \over \Gamma(s,\theta) + \kappa\,\rho(\theta)}.
\end{equation}  
Note that, for $\theta > \theta_0$, we can take the limits $q \to 0$ and $\theta \to \theta_0$ in different orders and get different answers.  If we take the limit of vanishing $q$ first, so that $\Gamma(s,\theta)$ vanishes, then $\chi \to \theta$.  However, if we let  $\theta$ go to $\theta_0$ first, so that $\rho(\theta)$ vanishes, then $\chi \to \hat\chi(q) \to \chi_0$ for sufficiently small $q$.  We see no {\it a priori} reason why the crossover between these limiting behaviors described by Eq.(\ref{chiss}) cannot be physically correct for a temperature independent $\chi_0$; but that assumption is qualitatively inconsistent with the HL data.  Equation (\ref{chiss}) implies that $\chi$ always must lie between $\theta$ and $\hat\chi(q)$; but we have found (in a calculation not shown here) that, if we choose $\chi_0 = \theta_0$ to be a $\theta$-independent constant, then the HL data in Fig.\ref{HL:qchi} for  $\theta = 0.31$ and $0.44$ lie above the allowed range in the transition region between the small-$q$ and large-$q$ limits.  

A physically plausible alternative is to assume first that $\chi_0 \ge \theta_0$ (the inequality being consistent with the existence of transient stress peaks), and then that $\hat\chi(q \to 0) = \theta$ whenever $\theta$ exceeds $\chi_0$.  That is, 
\begin{equation}
\label{hatchi0}
\lim_{q\to 0}\,\hat\chi(q) = \cases {\chi_0 & if $\theta < \chi_0$,\cr \theta &if $\theta > \chi_0$.}
\end{equation}
With this assumption, $\hat\chi(q \to 0)$ is a continuous function of $\theta$ across the glass transition and, for $\theta > \chi_0$, there are no small-$q$ steady-state solutions other than $\chi = \theta$. Nothing in this hypothesis precludes $\chi_0$ itself from being a function of temperature.  In fact, we think we see a hint in the HL data that $\chi_0$ may be a decreasing function of $\theta$ up to the point where $\theta = \chi_0$, beyond which $\hat\chi(q \to 0) = \theta$.  

\begin{figure}[h]
\centering \includegraphics[height=9.5 cm]{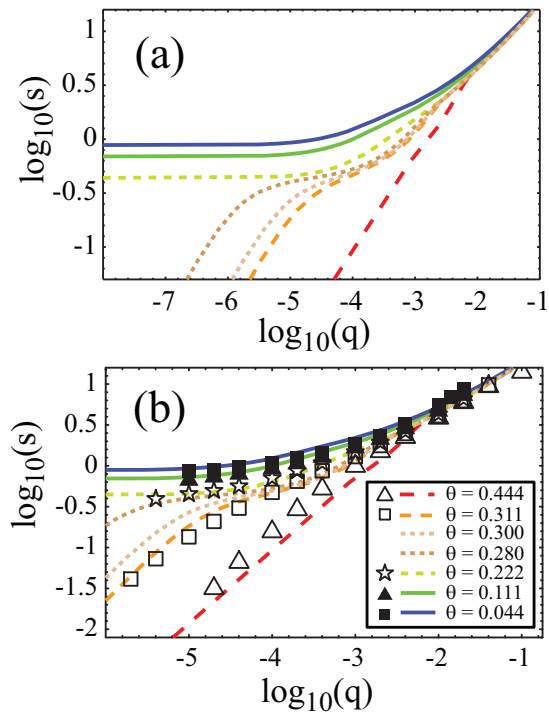}
\caption{\label{HL:sq} (a) Theoretical log-log plots of stress $s$ as a function of strain rate $q$ for a range of temperatures above and below the glass transition. The short-dashed curves are for $\theta = 0.28$ and $\theta=0.30$, both just above the glass transition temperature $\theta_0 = 0.2$.  (b) The same as (a) but for a smaller range of strain rates and including HL data at the selected temperatures}
\end{figure}

The assumption in Eq.(\ref{hatchi0}) accurately fits the HL data.  In implementing Eq.(\ref{hatchi0}), we have simply rescaled the parameters $\chi_A$ and $\chi_1$ in Eq.(\ref{alpha-chi}) so that the ratios $\chi_A/\chi_0$ and $\chi_1/\chi_0$ remain the same in both low and high temperature regions.  We also have chosen $\kappa = 5$ in Eq.(\ref{K-chi}) and $\tilde\mu = 0.3$ in Eq.(\ref{R-s}). Our results are shown in Figs. \ref{HL:qchi}, \ref{HL:qchiall}, \ref{HL:schi},  and \ref{HL:sq}. 

Figure \ref{HL:qchi} shows $\chi$ as a function of strain rate $q$ for a set of temperatures $\theta$ both above and below $\theta_0$.  As expected, the HL data and STZ curves collapse to roughly a single curve for temperatures below $\theta_0$.  Above $\theta_0$, as predicted by  Eq.(\ref{chiss}), $\chi \to \theta$ in the limit of small $q$; and the crossover to large-$q$ behavior is correctly predicted by the theory. Figure \ref{HL:qchiall} shows the same data as in Fig. \ref{HL:qchi} but here as the Arrhenius plot, Fig. \ref{Arr1}b, with the high-temperature data included.  We show this different view of the data to emphasize that even the high-temperature values of $\chi$ collapse to a single Arrhenius line, extrapolating to a single ``melting'' strain rate $q_0$, in the limit of infinite $\chi$.

Similarly, the graphs of stress $s$ as functions of $1/\chi$ in Fig. \ref{HL:schi}, and those for $s$ as functions of strain rate in Figs.~\ref{HL:sq}(a) and (b), show good agreement between theory and simulations at all temperatures and stresses.  As in the preceding figures, these curves and the corresponding HL data points are shown at a selection of temperatures above and below $\theta_0$. Note, in Fig.~\ref{HL:sq}, that the curves for $\theta > \theta_0$ make smooth transitions to linear viscous behavior at small strain rates. As expected, the curves for $\theta < \theta_0$ level off at temperature-dependent yield stresses. The short-dashed lines in Fig.~\ref{HL:sq}(a) illustrate the bifurcation at $\theta_0$.  At temperatures just above $\theta_0$, the material appears to exhibit a yield stress down to very small strain rates, but eventually flows even in the limit of arbitrarily small stress, thus showing how hard it is to make an accurate estimate of $\theta_0$ by going to smaller and smaller strain rates. 

Two material-specific parameters, $\tilde\mu$ and $s_1$, are best constrained using the HL data for $\log(s/q)$ vs. $\log(q)$, shown in Fig.~\ref{HL:etaq}. The small strain-rate asymptote of these curves for temperatures above the glass transition is $\log(\eta(\theta))$, and the way in which the stress crosses over from its power-law dependence at large $q$ to the yield stress or Newtonian viscosity at small $q$ is strongly sensitive to STZ transition rate parameters $\tilde\mu$ and $s_1$. We find that $\tilde\mu \cong 0.3$ and $s_1 \cong 0.08$ fit the crossover behavior quite well.

 The curves in Fig.~\ref{HL:etaq} show the STZ solutions for $\log(s/q)$ vs. $\log(q)$ at temperatures above and below the glass transition. For temperatures above the glass transition, $\hat{\eta}(q) = s/q$ is very similar to the Cross form used by HL to extrapolate their data.  However, the STZ solutions for $\hat{\eta}(q)$ asymptote to a slightly smaller value of $\eta$ as $q \to 0$ than the one specified by HL. It is unclear how to best extrapolate this stress/strain rate data to $s=0$, and this uncertainty explains why the HL viscosity values in Fig.~\ref{Arr1}a are slightly larger than the STZ fit.

\section{Concluding Remarks}

The natural way in which the STZ theory accounts for the wealth of data provided by the Haxton-Liu simulations lends credibility to major elements of that theory, especially the form of the expression for the plastic strain rate in Eq.(\ref{q-s}) and the central role played by the effective disorder temperature.  The self-consistent evaluation of parameters in Section \ref{HLdata} depends directly on the first of these elements; and the remarkable appearance of an effective glass transition in the nonequilibrum glassy state seems to confirm the second. Nevertheless, unanswered questions abound.  We conclude by listing some of them.

{\it How close is the analogy between the thermal and effective-thermal glass transitions?} The analogy does seem surprisingly close, but there are interesting and potentially very important differences.  We think that the near  equality of the transition temperatures $\theta_0$ and $\chi_0$, if correct at all, must be an artifact of the HL model. We have argued here that $\theta_0$ must be substantially smaller than $\chi_0$ in many realistic circumstances; and we suspect that this inequality might even turn out to be true in the HL model if measurements could be made at appreciably smaller strain rates.  It already is apparent that the effective Arrhenius activation energy $A\,k_B\,T_Z$ in Eq.(\ref{q-chi}) is larger than the thermal activation energy $k_B\,T_Z$ by a factor of about $1.5$; thus there is no universality in the energy scales.  Why not?  

We see an even deeper question in this regard.  As discussed in the paragraph preceding Eq.(\ref{q-chi}), the fact that $\hat\chi(q)$ goes to a nonzero value $\chi_0$ in the limit of small $q$ seems to have a robust explanation in purely kinetic terms.  Is a similar argument relevant to the thermal glass transition?  Would it imply that the viscosity -- the analog of $q^{-1}$ -- rigorously does diverge at $\theta_0$?  

{\it What should the STZ transition rate $R(s)$ really look like?  Which features of it are model dependent?  Which have some degree of universality?} In Eq.(\ref{R-s}), we have tried to go part of the way toward deducing $R(s)$ from first-principles, many-body physics.  Doing this correctly would be analogous to using statistical mechanics to compute the viscosity in the Navier Stokes equations.  The problem here is harder, of course, because we do not yet have a precise definition of an STZ. Nevertheless, this problem needs to be given more thought in view of the new information we have obtained from the HL results. 

One especially intriguing aspect of this question emerges from our observation that $R(s)$ exhibits Bagnold scaling at large $s$.  Under what circumstances may we expect this to happen?  Is it special to the HL model? Ordinarily, we think of Bagnold scaling as being relevant to the large-stress behavior of the strain rate as a whole, i.e. to $q(s)$ in present notation; whereas our $q(s)$ behaves quite differently.  The special feature here is that the Bagnold behavior is modified by the STZ density, i.e. the factor $\exp\,(-1/\chi)$ in Eq.(\ref{q-s}), which is strongly $q$ dependent.  We suspect that the full Bagnold behavior would occur at strain rates larger than $q_0$, where the STZ density is not a meaningful quantity, but where the HL model might look like a rapidly flowing granular material.  A better understanding of this large-$q$ situation might provide insight regarding the applicability of STZ theory to granular systems.

{\it Will the present version of the STZ theory accurately predict the time-dependent transient behavior of the HL model?}  One of the principal successes of earlier applications of the STZ theory has been that it correctly accounts for response times observed, for example, in transient stress-strain experiments.  The key idea here is that there is a separation of time scales between slow processes, i.e. relaxation of plastic flow or disorder temperature, as opposed to fast processes like the response to perturbations of the STZ orientation variable $m$ that appears in Eq.(\ref{m-s}).  That separation seems problematic at the upper end of the scale of strain rates used by HL.  We have no problem with it here because we consider only steady-state behavior.  

It will be  interesting to look at transient behavior in the HL model to see where the fully dynamic version of the STZ theory may break down.  Before doing that, however, it will be important to look at less esoteric behavior, e.g. transient responses to changes in driving forces of systems initially quenched at different rates to temperatures near the glass transition. Do such systems exhibit stress peaks?  The model that we have used for interpreting the HL data, in which $\chi_0 = \theta_0$, implies that they should not occur here. Might transient experiments reveal that, as we suspect, $\chi_0 > \theta_0$?  

{\it What are the implications for strain localization?} The effective temperature provides a mechanism for strain localization in amorphous materials -- regions with a higher effective temperature are more likely to undergo plastic deformation, which generates heat that increases the local effective temperature. Although previous studies assumed a constant value for $\hat{\chi}$~\cite{MANNING-LANGER}, Eq.~(\ref{q-chi}) shows that this assumption is only valid near the effective glass transition. The fact that $\hat{\chi}$ increases with increasing strain rate is likely to enhance localization and may permit steady state localized solutions. 

{\it What role do thermal fluctuations play in determining the properties of the effective glass transition? }  We already have concluded that $\hat\chi(q \to 0)$ must be equal to $\theta$ at high temperatures, $\theta > \chi_0$; but we have little knowledge about what happens to the corresponding $\chi_0$ when $\theta < \chi_0$, and have simply assumed that it is a constant.  More detailed information about this quantity might help us understand the molecular mechanisms that are at work here.

\begin{acknowledgments}
We thank T. Haxton and A. Liu for allowing us to use their data and for extensive discussions about this project.  M.L.M. acknowledges an NSF Graduate Research Fellowship and support from James S. McDonnell Foundation, the David and Lucile Packard Foundation, and NSF grant number DMR-0606092. J.S.L. was supported by DOE grant number DE-FG03-99ER45762.
\end{acknowledgments}

\end{document}